\documentclass[journal]{IEEEtran}

\usepackage{booktabs}
\usepackage{tabularx}

\usepackage{amssymb}
\usepackage{amsbsy}
\usepackage{amsmath}
\usepackage{bm}
\usepackage{verbatim}
\usepackage{cite}
\usepackage{mathrsfs}
\usepackage{amsfonts}
\usepackage{graphicx}
\usepackage[tight,footnotesize]{subfigure}
\usepackage[10pt]{moresize}
\usepackage{array}
\usepackage{color}
\usepackage{epsfig}
\usepackage{stfloats}
\usepackage{balance}

\usepackage{hyperref} 

\usepackage{amssymb}

\usepackage[noend]{algpseudocode}

\usepackage{algorithmicx,algorithm}

\usepackage{cite}

\usepackage{setspace}
\usepackage{cases}

\usepackage{graphicx}
\usepackage{epstopdf}
\usepackage{multirow}
\usepackage{extarrows}
\newcommand{\subparagraph}{}
\usepackage{titlesec}
\titlespacing{\section}{0pt}{1 ex plus .0ex minus .0ex}{1ex plus .0ex}
\titlespacing{\subsection}{0pt}{1.5 ex plus .0ex minus .0ex}{0.0 ex plus 0.0ex}
\titlespacing{\subsubsection}{0pt}{0.5ex plus .0ex minus .0ex}{0.0ex plus .0ex}

\usepackage{amsmath} 
\allowdisplaybreaks[4]

\ifCLASSINFOpdf

\else

\fi

\begin{document}
\title{A Graph-based Hybrid Beamforming Framework for MIMO Cell-Free ISAC Networks}

\author{Yanan~Du,~\IEEEmembership{Member,~IEEE,}
{Sai~Xu,~\IEEEmembership{Member,~IEEE,}
and~Jagmohan Chauhan

\vspace{-3mm}

\thanks{This work of Y. Du is supported by the European Research Executive Agency's Horizon Europe MSCA 2022 Postdoctoral Fellowship CIRED under Grant 101109336.}

\thanks{This work of S. Xu and J. Chauhan is supported by the Engineering and Physical Sciences Research Council (EPSRC) under Grant EP/X01200X/1.}

\thanks{Y.~Du is with the Department of Electronic and Electrical Engineering, University of Sheffield, Sheffield, S1 4ET, UK (e-mail: $\rm yanan.du@sheffield.ac.uk$).}

\thanks{S. Xu and J. Chauhan are with University College London, London, UK (e-mail: $\rm sai.xu, jagmohan.chauhan@ucl.ac.uk$).}

}
\thanks{Manuscript received XX XX, XXXX; revised XX XX, XXXX. }}
\maketitle
\begin{abstract}
This paper develops a graph-based hybrid beamforming framework for multiple-input multiple-output (MIMO) cell-free integrated sensing and communication (ISAC) networks. Specifically, we construct a novel MIMO cell-free ISAC network model. In this model, multiple dual-function base station (BS) transmitters employ distributed hybrid beamforming to enable simultaneous communication and sensing, while maintaining physical separation between the transmitters and the radar receiver. Building on this model, we formulate a multi-objective optimization problem under a power constraint to jointly improve communication and sensing performance. To solve it, the problem is first reformulated as a single-objective optimization problem. 
Then, a graph-based method composed of multiple graph neural networks (GNNs) is developed to realize hybrid beamforming with either perfect or imperfect channel state information. Once trained, the neural network model can be deployed distributively across BSs, enabling fast and efficient inference. To further reduce inference latency, a custom field-programmable gate array (FPGA)-based accelerator is developed. Numerical simulations validate the communication and sensing capabilities of the proposed optimization approach, while experimental evaluations demonstrate remarkable performance gains of FPGA-based acceleration in GNN inference.
\end{abstract}
\IEEEpeerreviewmaketitle
\section{Introduction}
%
\IEEEPARstart{C}{ell}-free network architectures are widely recognized as a transformative paradigm in wireless communications~\cite{Wang2023VisionsRequirements,Zhang2024InterdependentCell}. Unlike traditional cellular systems, which rely on geographically partitioned cells and dedicated base stations (BSs), these networks deploy numerous spatially dispersed access points (APs) coordinated by a centralized processing node~\cite{Chen2024Distributed}. By leveraging coherent joint processing and centralized coordination among APs, cell-free systems eliminate cell boundaries, enable cooperative transmission, suppress inter-cell interference, and extend coverage. They also enhance reliability and user experience in dense networks. Furthermore, their capability to support high user density and stringent latency requirements makes them particularly suitable for future wireless networks, which demand improved spectral efficiency, massive connectivity, and robust service continuity~\cite{Zhang2021JointPrecoding}. \par 
%
To support high-precision localization, intelligent applications, efficient spectrum utilization, and other requirements in next-generation wireless networks, mobile communication networks are increasingly considering the integration of sensing capabilities~\cite{Cui2021IntegatingSensing, Nasir2024JointUsers}. Integrated sensing and communication (ISAC) research can typically be categorized into two types: radar-communication coexistence (RCC)~\cite{Du2025Intelligent, Du2025uplink}, which emphasizes interference mitigation, allowing radar and communication systems to function simultaneously, and dual-function radar-communication (DFRC)~\cite{Xu2024Intelligent, Zhang2021AnOverview, Su2021SecureRadar}, which leverages a unified signal and shared hardware to jointly realize both functions. Compared with RCC, DFRC achieves higher integration, facilitates cooperation, and alleviates spectrum congestion~\cite{Liu2020JointRadar, Liu2023IntegratedSensing}. Building on this foundation, recent research has increasingly explored ISAC in cellular systems, including cell-free architectures~\cite{Demirhan2025Joint, Ren2025Secure, Cao2023JointResource, Mao2024CommunicationSensing, Salem2024IntegratedCooperative,Cao2023DesignandPerformance}, to leverage integration benefits and further enhance joint communication and sensing capabilities. \par
%
In recent years, cell-free ISAC networks have emerged as a key research direction. Demirhan \textit{et al.}~\cite{Demirhan2025Joint} studied cell-free ISAC multiple-input multiple-output (MIMO) networks and developed a joint beamforming scheme that balances sensing and communication performance, offering gains over conventional methods. Ren \textit{et al.}~\cite{Ren2025Secure} presented a secure joint beamforming approach for cell-free ISAC networks, countered both information and sensing eavesdropping, and attained optimality through semidefinite relaxation. Mao \textit{et al.}~\cite{Mao2024CommunicationSensing} examined a cell-free massive MIMO ISAC architecture, studying how target location ambiguity affects beamforming performance in both uplink and downlink channels. Salem \textit{et al.}~\cite{Salem2024IntegratedCooperative} investigated full-duplex cell-free MIMO ISAC systems leveraging reconfigurable intelligent surfaces (RIS) and developed a joint optimization framework to maximize weighted radar and communication signal-to-interference-plus-noise ratios (SINRs). Cao \textit{et al.}~\cite{Cao2023DesignandPerformance} proposed a cell-free massive MIMO architecture integrating communication and radar, employing vector orthogonal frequency-division multiplexing (OFDM) waveforms to enhance both communication quality and target detection reliability. Furthermore, Cao \textit{et al.}~\cite{Cao2023JointResource} extended their work to a user-centric cell-free massive MIMO ISAC system, where a low-complexity method was proposed to
jointly manage user scheduling and power distribution to optimize sum-rate performance.  Zhang \textit{et al.}~\cite{Zhang2024Integrated}  developed a tensor-based unified approach for massive MIMO-ISAC systems, enabling simultaneous estimation of channel and target parameters with enhanced sensing resolution and reduced training overhead.\par
On the other hand, thanks to its ability to lower hardware cost as well as power consumption while balancing communication and sensing performance, hybrid beamforming has attracted considerable research attention. Wang \textit{et al.}~\cite{Wang2022PartiallyConnected} proposed a Cramér-Rao bound-based multi-user hybrid beamforming design framework for ISAC systems, jointly optimizing analog and digital beamformers to improve the estimation accuracy of direction-of-arrival (DOA) while meeting the communication SINR constraints. Qi \textit{et al.}~\cite{Qi2022HybridBeamforming} investigated hybrid beamforming for mmWave MIMO ISAC systems, and transmit beams and phase vectors for DFRC BSs were optimized alternately while ensuring compliance with SINR and power constraints.
Leyva \textit{et al.}~\cite{Leyva2024HybridBeamforming} proposed a fully-connected hybrid beamforming method for multi-beam, multi-user ISAC, which used iterative alternate optimization to achieve weighted sum-rate maximization under power and sensing restrictions, achieving near fully digital performance and outperforming existing methods. Wang \textit{et al.}~\cite{Wang2025JointHybrid} investigated millimeter-wave OFDM ISAC systems with RIS, jointly designing hybrid beamforming as well as phase shifts to enhance sensing and communication performance under SINR and power constraints. Li \textit{et al.}~\cite{Li2025SecureHybrid} proposed secure hybrid beamforming for millimeter-wave ISAC systems using subarray architectures, designing dual-functional signals and dynamic subarrays to enhance sensing and secrecy performance under imperfect channel state information (CSI).\par
Building on existing research, we propose a graph-based hybrid beamforming framework for MIMO cell-free ISAC networks. In contrast to prior studies, the key distinctions of this work lie in the network architecture, algorithm design, and hardware implementation. Specifically, we adopt a transceiver-separation design with hybrid beamforming at the transmitters. On the optimization algorithm side, a graph-based method composed of multiple graph neural networks (GNNs) is developed for distributed DFRC design, enabling effective coordination across multiple access points. Furthermore, a customized  field-programmable gate array (FPGA)-based accelerator is developed to accelerate the inference process for the unique GNN architecture and ensure real-time execution, thereby significantly enhancing the system’s practical deployment potential. Specifically, this paper makes the following main contributions: \par
\begin{itemize}
\item We build a novel MIMO cell-free ISAC network architecture, where the DFRC transmitters at BSs employ distributed hybrid beamforming to concurrently execute wireless communication and radar sensing, while the transmitters and radar receivers are physically separated. Based on this model, a multi-objective optimization framework is constructed under the imposed power constraints, with the objective of jointly enhancing the performance of both communication and sensing. \par
\item A graph-based methodology is introduced for DFRC design, in which the multi-objective optimization framework is transformed into a single-objective formulation. Based on this formulation, we introduce a hybrid beamforming scheme based on multiple-GNN, where both communication and sensing channel data are taken as inputs, and the digital and analog beamforming are produced as outputs. After centralized training, the neural network can be deployed in a distributed manner across BSs, thereby enabling rapid and efficient inference.
\item  We designed a custom FPGA-based accelerator specifically tailored to the unique GNN architecture for DFRC design. In contrast to existing FPGA-based GNN accelerators, which require CPU control and external memory access due to incompatibility with our model, the proposed design eliminates these overheads, thereby enabling efficient, low-latency processing well-suited for real-time deployment in MIMO cell-free ISAC networks.
\item Numerical simulations have been conducted to assess the communication and sensing capabilities of the  MIMO cell-free ISAC network, confirming the superiority of the hybrid beamforming and graph-based approach. Additionally, experimental results evaluate the effectiveness of the FPGA-based accelerator in enhancing the speed and efficiency of GNN inference.
\end{itemize}
%

%
%
\par The rest of this paper is structured as follows. In section II, a MIMO cell-free ISAC network is modeled and then a maximization problem for optimizing communication and sensing performance is formulated. Section III introduces a graph-based optimization algorithm that achieves satisfactory communication and sensing performance with reduced computational complexity. Section IV details the FPGA-based accelerator designed to reduce GNN inference latency. Simulation and experimental results in Section V demonstrate the communication and sensing performance achieved by the proposed schemes as well as the computational performance of the FPGA-based accelerator. Section VI concludes the paper by summarizing the primary outcomes.

\section{System Model and Problem Formulation}
\begin{figure}
\centering
\includegraphics[width= 3.5 in]{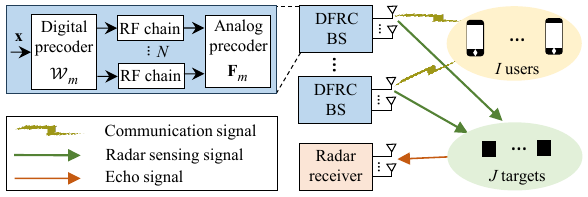}
\caption{An illustration of MIMO cell-free ISAC network.}
\label{Fig1}
\end{figure}
\subsection{System Model}
\begin{figure*}[b]
\hrulefill
\setcounter{equation}{0}
\begin{align}
\textbf{y}_{i}&=  \sum_{m \in \mathcal{M}} \textbf{H}_{\text{com},m, i}\textbf{F}_m \textbf{w}_{\text{com},m,i} x_i + \underbrace{ \sum_{i^{\prime} \in \mathcal{I}, i^{\prime}  \neq i } \sum_{m \in \mathcal{M}} \textbf{H}_{\text{com},m, i}\textbf{F}_m \textbf{w}_{\text{com},m,i'} x_{i'} +  \sum_{j \in \mathcal{J}} \sum_{m \in \mathcal{M}}  \textbf{H}_{\text{com},m, i}\textbf{F}_m \textbf{w}_{\text{sen},m, j}}_\text{Interference signals}  +\textbf{n}_{i},  \label{1} \\
\textbf{y}_{j}&= \sum_{m \in \mathcal{M}} \textbf{H}_{\text{sen},m, j}\textbf{F}_m \textbf{w}_{\text{sen},m,j}+\underbrace{\sum_{j^{\prime} \in \mathcal{J}, j^{\prime}  \neq j } \sum_{m \in \mathcal{M}}\textbf{H}_{\text{sen},m, j}\textbf{F}_m \textbf{w}_{\text{sen},m,j^{\prime}}+\sum_{i \in \mathcal{I} } \sum_{m \in \mathcal{M}}\textbf{H}_{\text{sen},m, j}\textbf{F}_m \textbf{w}_{\text{com},m,i}x_i}_\text{Interference signals}+\textbf{n}_\text{r}, \label{2} \\
\bar{\textbf{y}}_{j}&= \sum_{m \in \mathcal{M}}\boldsymbol{\textbf{u}}_{j} \textbf{H}_{\text{sen},m, j}\textbf{F}_m \textbf{w}_{\text{sen},m,j}+\sum_{j^{\prime} \in \mathcal{J}, j^{\prime}  \neq j } \sum_{m \in \mathcal{M}}\boldsymbol{\textbf{u}}_{j} \textbf{H}_{\text{sen},m, j}\textbf{F}_m \textbf{w}_{\text{sen},m,j^{\prime}}   +\sum_{i \in \mathcal{I} } \sum_{m \in \mathcal{M}}\boldsymbol{\textbf{u}}_{j}\textbf{H}_{\text{sen},m, j}\textbf{F}_m\textbf{w}_{\text{com},m,i}x_i+\boldsymbol{\textbf{u}}_{j}\textbf{n}_\text{r}, \label{3} \\
\gamma_{i} &= \left(\sum_{m \in \mathcal{M}}\textbf{H}_{\text{com},m, i}\textbf{F}_m \textbf{w}_{\text{com},m,i}\right)^{H} \Bigg( \sigma_{i}^2 \textbf{I}_{i}   +\sum_{i^{\prime} \in \mathcal{I}, i^{\prime}  \neq i }  ( \sum_{m \in \mathcal{M}}\textbf{H}_{\text{com},m, i}\textbf{F}_m \textbf{w}_{\text{com},m,i'})(\sum_{m \in \mathcal{M}}\textbf{H}_{\text{com},m, i} \textbf{F}_{m}\textbf{w}_{\text{com},m,i'})^H   \nonumber \\
&~~ +\sum_{j \in \mathcal{J}}(\sum_{m \in \mathcal{M}} \textbf{H}_{\text{com},m, i}\textbf{F}_{m} \textbf{w}_{\text{sen},m, j})( \sum_{m \in \mathcal{M}}\textbf{H}_{\text{com},m, i}\textbf{F}_{m} \textbf{w}_{\text{sen},m, j})^H      \Bigg) ^{-1} \left(\sum_{m \in \mathcal{M}}\textbf{H}_{\text{com},m, i}\textbf{F}_{m}\textbf{w}_{\text{com},m,i} \right), \label{4} \\
\gamma_{j} &= \frac{|\sum_{m \in \mathcal{M}}\boldsymbol{\textbf{u}}_{j}\textbf{H}_{\text{sen},m, j}\textbf{F}_{m}\textbf{w}_{\text{sen},m,j}|^2}{\sigma_\text{r}^2 |\boldsymbol{\textbf{u}}_{j}|^2 +  \sum_{j^{\prime} \in \mathcal{J}, j^{\prime}  \neq j } |\sum_{m \in \mathcal{M}}\boldsymbol{\textbf{u}}_{j}\textbf{H}_{\text{sen},m,j}\textbf{F}_{m}\textbf{w}_{\text{sen},m,j^{\prime}}|^{2} +  \sum_{i \in \mathcal{I} } |\sum_{m \in \mathcal{M}}\boldsymbol{\textbf{u}}_{j}\textbf{H}_{\text{sen},m, j}\textbf{F}_{m}\textbf{w}_{\text{com},m,i}|^{2}}.  \label{5}
\end{align}
\vspace*{0pt}
\end{figure*}
Consider a MIMO cell-free ISAC network composed of $M$ DFRC MIMO BSs, a dedicated radar receiver for collecting sensing signal echo, $I$ communication  users and $J$ targets, as illustrated in Fig. \ref{Fig1}. Each BS is outfitted with $N$ transmit radio frequency (RF) chains and $N_\text{t}$ antennas, with $I+J\leq N \leq N_\text{t}$ assumed. Each communication user and the radar receiver possess $N_\text{u}$ and $N_\text{r}$ antennas, respectively. All the BSs employ fully connected hybrid beamforming architectures with phase shifter-based analog precoder set $\mathcal{F}=\{\textbf{F}_{1}, ...,\textbf{F}_{m}, ..., \textbf{F}_{M} \}$ and digital precoder set $\mathcal{W}=\{\mathcal{W}_1, ..., \mathcal{W}_m, ..., \mathcal{W}_M \}$ with $\mathcal{W}_m=\{\textbf{w}_{\text{com},m,1}, ..., \textbf{w}_{\text{com},m,I}, \textbf{w}_{\text{sen},m,1}, ..., \textbf{w}_{\text{sen},m,J} \}$, where $\textbf{F}_{m}\in \mathbb{C}^{N_\textbf{t} \times N}$ denotes the analog precoding matrix at the $m$-th BS, and $\textbf{w}_{\text{com},m,i}\in \mathbb{C}^{N \times 1}$ and $\textbf{w}_{\text{sen},m,j}\in \mathbb{C}^{N \times 1}$ represent the digital beamforming vectors at the $m$-th BS for the $i$-th user and the $j$-th target, respectively. Different from traditional DFRC MIMO BSs, the considered BSs are only responsible for transmitting both communication and sensing signals simultaneously, without collecting the sensing echoes. Instead, a separate radar receiver is employed for sensing signal reception. Since all BSs and the radar receiver are interconnected via high-speed optical fiber, the latency in their information exchange can be regarded as negligible. Moreover, both the communication and sensing channels are considered to vary slowly.\par
Let $\mathcal{M}=\{1,2,...,M\}$, $\mathcal{N}=\{1,2,...,N\}$, $\mathcal{I}=\{1,2,...,I\}$ and $\mathcal{J}=\{1,2,...,J\}$ denote the index sets of BSs, RF chains per BS, users, and sensing targets, respectively. In the considered MIMO cell-free ISAC network architecture, the same data stream is delivered to each user by all the BSs. The channels for communication and sensing are represented as Rician fading processes with varying Rician factors, where the Rayleigh channel can be regarded as a special instance without the line-of-sight (LoS) component. The antenna arrays at the BSs, users, and radar receiver are assumed to be linear. Accordingly, the dual-functional transmitted signal at the $m$-th BS is expressed as $\sum_{i \in \mathcal{I}} \textbf{F}_{ m}\textbf{w}_{\text{com},m,i} x_i +  \sum_{j \in \mathcal{J}} \textbf{F}_{m} \textbf{w}_{\text{sen},m,j}$, where $\textbf{x} = [x_1, x_2, \ldots, x_I]^T$ denotes the vector of data streams, and $x_i$ represents the data symbol intended for the $i$-th user. \par
Following the system model, the received signal at the $i$-th user and the radar-received signal from the $j$-th target are given by \eqref{1} and \eqref{2}, respectively. Here, $\textbf{n}_{i}\sim \mathcal{CN}(\textbf{0},\sigma_{i}^{2} \textbf{I}_{i})$ and $\textbf{n}_\text{r} \sim \mathcal{CN}(\textbf{0},\sigma_\text{r}^{2} \textbf{I})$ represent the additive white Gaussian noise at the $i$-th user and the radar receiver, respectively. The matrices  $\textbf{H}_{\text{com},m,i}\in \mathbb{C}^{N_\text{u} \times N_\text{t}}$ and $\textbf{H}_{\text{sen},m,j}\in \mathbb{C}^{N_\text{r} \times N_\text{t}}$ represent the communication channel from the $m$-th BS to the $i$-th user and the sensing channel from the $m$-th BS to the radar receiver via the $j$-th target, respectively, and are mathematically expressed as
\begin{align}
\textbf{H}_{\text{com},m,i} &= \text{diag}(\textbf{b}_{m,i}) \textbf{H}_{m,i} \text{diag}(\textbf{a}_{m,i}) , \nonumber \\
\textbf{H}_{\text{sen},m,j} &= \text{diag}(\textbf{g}_{j}) \textbf{H}_{m,j} \text{diag}(\textbf{c}_{m,j}), \nonumber
\end{align}
where $\textbf{a}_{m,i}$ and $\textbf{b}_{m,i}$ denote the steering vectors associated with the azimuth angle $\theta_{m,i}$ from the $m$-th BS to the $i$-th user. Similarly, $\textbf{c}_{m,j}$  represents the steering vector corresponding to the azimuth angle  $\theta_{m,j}$ from the $m$-th BS to the $j$-th target, while $\textbf{g}_{j}$ corresponds to the azimuth angle $\theta_{j}$ from the $j$-th target to the radar receiver. Their mathematical expressions are given by
\begin{align}
\textbf{a}_{m,i}&= \frac{1}{N_\text{t}} [1, e^{j 2 \pi d_{m} \text{sin}(\theta_{m,i})},...,e^{j 2 \pi (N_\text{t}-1) d_{m} \text{sin}(\theta_{m,i})}]^{T}, \nonumber \\
\textbf{b}_{m,i} &= \frac{1}{N_\text{u}} [1, e^{j 2 \pi d_i \text{sin}(\theta_{m,i})},...,e^{j 2 \pi (N_\text{u}-1) d_i \text{sin}(\theta_{m,i})}]^{T},  \nonumber \\
\textbf{c}_{m,j}&= \frac{1}{N_\text{t}} [1, e^{j 2 \pi d_{m} \text{sin}(\theta_{m,j})},...,e^{j 2 \pi (N_\text{t}-1) d_{m} \text{sin}(\theta_{m,j})}]^{T}, \nonumber \\
\textbf{g}_{j} &= \frac{1}{N_\text{r}} [1, e^{j 2 \pi d_\text{r} \text{sin}(\theta_{j})},...,e^{j 2 \pi (N_\text{r}-1) d_\text{r} \text{sin}(\theta_{j})}]^{T},  \nonumber
\end{align}
where $d_{m}$, $d_{i}$ and $d_\text{r}$ are the antenna spacings, normalized by wavelength, at the $m$-th BS, the $i$-th user and the radar receiver, respectively. By applying the normalized receive beamforming vector $\boldsymbol{\mathbf{u}}_{j}$ at the radar receiver for the echo from the $j$-th target, the resulting output is given by \eqref{3}. Subsequently, the SINRs for the $i$-th user and the radar receiver corresponding to the $j$-th target are given by \eqref{4} and \eqref{5}, respectively.\par
\subsection{Problem Formulation}
This research focuses on improving the communication and sensing capabilities of the considered network, which is cast into a multi-objective optimization problem as follows:
\begin{align*}
(\text{P}0)~  \text{Q}1 :& \mathop{\text{max}}\limits_{\textbf{F}_{m}, \textbf{w}_{\text{com},m,i}} \sum_{i \in \mathcal{I}} \text{log}(1+\gamma_i) \\
\text{Q}2 :& \mathop{\text{max}}\limits_{\textbf{F}_{m}, \textbf{w}_{\text{sen},m,j}} \sum_{j \in \mathcal{J}} \text{log}(1+\eta\gamma_j) \\
  \text{s.t.}  ~ &  \text{C1}: \text{Tr}\Bigg(\sum_{i \in \mathcal{I}} \textbf{F}_m\textbf{w}_{\text{com},m,i}\textbf{w}_{\text{com},m,i}^H\textbf{F}_m^H \\ & ~~~~ +  \sum_{j \in \mathcal{J}} \textbf{F}_m \textbf{w}_{\text{sen},m,j}\textbf{w}_{\text{sen},m,j}^H\textbf{F}_m^H \Bigg) \leq P, ~m \in \mathcal{M}, \\
   & \text{C2}: |\textbf{F}_m (n_\text{t},n)|^2=1,  ~m \in \mathcal{M}, n_\text{t}\in \mathcal{N_\text{t}},n \in \mathcal{N}, 
\end{align*}
where $P$ represents the overall power budget allocated to each BS. $\text{log}(1+\eta \gamma_j)$ is defined to quantify the radar sensing capacity, where the scaling factor $\eta$ is employed to mitigate the significant magnitude disparity between the sensing and communication SINRs, thereby enabling a more balanced compromise between sensing and communication performance. $\mathcal{N_\text{t}}$ represents the antenna set at each BS. Constraint C1 ensures that the transmit power of each BS is constrained by the given budget, whereas C2 guarantees the constant-modulus characteristic of the phase shifters within the analog beamforming matrix $\mathbf{F}_m$. It should be emphasized that the radar sensing capacity does not carry direct physical meaning; rather, it acts as a surrogate metric that indirectly indicates the effectiveness of radar sensing.\par
By applying the weighted sum method \cite{Marler2004Surveyofmulti}, the problem (P0) is transformed into
\begin{align*}
(\text{P}1)~~ \mathop{\text{max}}\limits_{\textbf{F}_m, \textbf{w}_{\text{com},m,i},\textbf{w}_{\text{sen},m,j}}      & \alpha_\text{com}\sum_{i \in \mathcal{I}} \text{log}(1+\gamma_i) \\
& \quad\quad\quad\quad+  \alpha_\text{sen}\sum_{j \in \mathcal{J}} \text{log}(1+ \eta \gamma_j),\\
  \text{s.t.}  \quad  & \text{C1} ~ \text{and}~ \text{C2}, 
\end{align*}
where $\alpha_\text{com}$ and $\alpha_\text{sen}$ denote the weighting coefficients used to balance the Pareto trade-off between communication and sensing. The resulting objective function is termed the weighted sum communication and sensing capacity (WSCSC).
\section{GNN-Based Optimization}
\begin{figure*}
\centering
\includegraphics[width= 6.6 in]{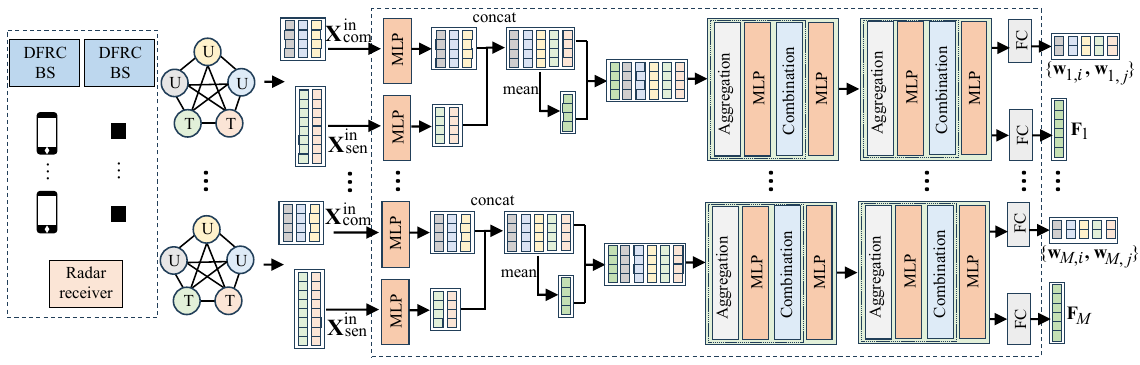}
\caption{Graph representation and neural network architecture.}
\label{Fig2}
\end{figure*}

In this section, a graph-based optimization framework is presented to efficiently obtain near-optimal solutions for the non-convex problem (P1). To apply the graph-based approach, the MIMO cell-free ISAC network is first represented as a graph, which serves as the input to the GNN. Within this framework, a multi-GNN architecture is employed, as illustrated in Fig.~\ref{Fig2}, where all GNNs possess a homogeneous structure and operate under the same principles.
\subsection{Input}
The initial layer of each GNN is built using two multi-layer perceptron (MLP) layers. The input to this layer is derived from a graph representation $\mathcal{G} = (\mathcal{V}, \mathcal{E})$, where $\mathcal{V}$ and $\mathcal{E}$ denote the nodes and edges of the network, respectively. For the optimization problem (P1), the interdependence between communication and sensing channels makes it natural to encode these as node-level features. Given that each BS has $N_\mathbf{t}$ antennas, the MLP input dimension is set to $2N_\mathbf{t}$ to match the total number of node features. The resulting node feature matrices are formally defined as follows:
\begin{equation*}
\textbf{X}_{\text{com},m}^{\text{in}} =
\begin{bmatrix}
\text{Re}\{\textbf{H}_{\text{com},m,1}\} & \text{Im}\{\textbf{H}_{\text{com},m,1}\}   \\
\vdots & \vdots  \\
\text{Re}\{\textbf{H}_{\text{com},m,i}\} & \text{Im}\{\textbf{H}_{\text{com},m,i}\} \\
\vdots & \vdots  \\
\text{Re}\{\textbf{H}_{\text{com},m,I}\} & \text{Im}\{\textbf{H}_{\text{com},m,I}\} 
\end{bmatrix},  \nonumber
\end{equation*}
\begin{equation*}
\textbf{X}_{\text{sen},m}^{\text{in}} =
\begin{bmatrix}
\text{Re}\{\textbf{H}_{\text{sen},m,1}\} & \text{Im}\{\textbf{H}_{\text{sen},m,1}\}     \\
\vdots & \vdots  \\
\text{Re}\{\textbf{H}_{\text{sen},m,j}\} & \text{Im}\{\textbf{H}_{\text{sen},m,j}\} \\
\vdots & \vdots  \\
\text{Re}\{\textbf{H}_{\text{sen},m,J}\} & \text{Im}\{\textbf{H}_{\text{sen},m,J}\} 
\end{bmatrix}.\nonumber
\end{equation*}
where $\textbf{H}_{\text{com},m}$ and $\textbf{H}_{\text{sen},m}$ represent the averaged channel matrices over all users ($\textbf{H}_{\text{com},m,i}$ for $i \in \mathcal{I}$) and targets ($\textbf{H}_{\text{sen},m,j}$ for $j \in \mathcal{J}$), respectively. Notably, the channel vectors corresponding to different users and targets are fed into the MLP layers simultaneously in parallel. Passing through the MLP layer produces the intermediate output representation vectors $\textbf{z}_\text{com}^\text{1}$ and $\textbf{z}_\text{sen}^\text{1}$. These intermediate representations can be formally expressed as
\begin{equation*}
\textbf{z}_\text{com}^\text{1} = f_\text{MLP} (\textbf{z}_{\text{com}}^{\text{in}}), \nonumber  ~
\textbf{z}_\text{sen}^\text{1} = f_\text{MLP} (\textbf{z}_{\text{sen}}^{\text{in}}). \nonumber
\end{equation*}
In this context, $f_\textsc{MLP}(\cdot)$  denotes the transformation implemented by the MLP layer, which consists of two fully connected (FC) layers employing an identical activation function. The intermediate representation vectors $\textbf{z}_\text{com}^\text{1}$ and $\textbf{z}_\text{sen}^\text{1}$ are then combined, which can be mathematically described as
\begin{equation*}
\textbf{z}^\text{1} = f_\text{concat} (\textbf{z}_\text{com}^\text{1},\textbf{z}_\text{sen}^\text{1}). \nonumber  
\end{equation*}
where $f_\text{concat}(\cdot)$ denotes the concatenation operation. Since the design of the analog beamforming matrix $\textbf{F}_{m}$ depends on both $\textbf{H}_{\text{com},m, i}$ and $\textbf{H}_{\text{sen},m, j}$, the feature vector $\textbf{z}^\text{1}$ is first averaged to obtain $\textbf{z}_\text{mean}^\text{1}$. Then, $\textbf{z}^\text{1}$ and $\textbf{z}_\text{mean}^\text{1}$ are concatenated along the feature dimension to form a joint representation, which serves as the input for generating the beamforming matrix, expressed mathematically as
\begin{equation*}
\textbf{z}^\text{2} = f_\text{concat} (\textbf{z}^\text{1},\textbf{z}_\text{mean}^\text{1}) \nonumber  
\end{equation*}
Here, $\textbf{z}_\text{mean}^\text{1} = \left[ \text{Re}\{\textbf{H}_{m}\}, \text{Im}\{\textbf{H}_{m}\}\right]$, where   $\text{Re}\{\textbf{H}_{m}\}$ denotes the mean of all $\text{Re}\{\textbf{H}_{\text{com},m,i}\}$ and $\text{Re}\{\textbf{H}_{\text{sen},m,j}\}$ for $i \in \mathcal{I}$ and $j \in \mathcal{J}$, and $\text{Im}\{\textbf{H}_{m}\}$ denotes the mean of all $\text{Im}\{\textbf{H}_{\text{com},m,i}\}$ and $\text{Im}\{\textbf{H}_{\text{sen},m,j}\}$ for $i \in \mathcal{I}$ and $j \in \mathcal{J}$.
\subsection{Graph Convolution Module Design}
In the proposed GNN framework, the core component is a two-layer graph convolution module. Both convolutional layers adopt the same structural design and computational scheme. For clarity, we consider the first convolutional layer as an example for illustration. \par
Let $\mathbf{z}_{m}^{2}$ denote the input matrix to the convolutional layer, where $m$ indexes the $m$-th BS. The vector $\mathbf{z}_{m, k}^{2}$ represents the $k$-th row of $\mathbf{z}_{m}^{2}$, where $k \in \{1, 2, \dots, I + J + 1\}$. Here, $I$ and $J$ denote the numbers of communication and sensing channels, respectively, and the additional row corresponds to the average of the communication and sensing channels. The processing procedure is as follows. 
First, $z_{m,k}^{(2)}$ is passed through an initial MLP. Then, feature representations from neighboring channels $\{ z_{m,k'}^{(2)} \}_{k' \in \mathcal{K} \setminus k}$ are aggregated through an aggregation function $f_\text{agg}(\cdot)$. This aggregated neighborhood information is then combined with the original node feature via a combination function $f_\text{com}(\cdot)$, which is followed by another MLP to yield the updated representation $z_{m,k}^{(3)}$. The complete transformation can be formally expressed as:
\begin{align}
z_{m,k}^{(3)} = f_\text{com} \left(z_{m,k}^{(2)}, f_\text{agg} \left( \{ z_{m,k'}^{(2)} \}_{k' \in \mathcal{K} \setminus k} \right) \right), \nonumber
\end{align}
where $f_\text{com}(\cdot)$ denotes a feature combination operation implemented as a concatenation followed by an MLP. The aggregation function $f_\text{agg}(\cdot)$ is defined as:
\begin{align}
f_\text{agg} \left( \{ z_{m,k'}^{(2)} \}_{k' \in \mathcal{K} \setminus k} \right) = \psi \left( \{ f_\text{MLP} (z_{m,k'}^{(2)} ) \}_{k' \in \mathcal{K} \setminus k} \right), \nonumber
\end{align}
where $f_\text{MLP}(\cdot)$ is a shared MLP applied to each neighbor’s feature, and $\psi(\cdot)$ denotes an element-wise max-pooling function that is invariant to permutations of its input. This ensures that the aggregation process is order-independent with respect to the neighboring nodes. \par
In this context, a nested aggregation function, combining an MLP and element-wise max-pooling, is employed to effectively capture and merge feature information from the neighboring nodes. The combination function further integrates this aggregated neighborhood feature with the node’s own representation, allowing the network to learn the interaction between the user and target channels within each BS. Both convolutional layers in the GNN follow this design to capture these interactions and enhance interference mitigation. Importantly, the proposed GNN-based approach is inherently scalable and generalizable to varying numbers of users and targets, since all user and target channels are treated as node features and merged into a unified tensor that serves as the network input.
\subsection{Output}
The neural network is designed to yield two distinct output tensors. For the $m$-th GNN, the output first passes through a FC layer, where the mapping is applied from the first row up to the $(I+J)$-th row. This branch produces the digital beamforming vectors $\textbf{w}_{\text{com},m,i}$ for $i \in \mathcal{I}$ and $\textbf{w}_{\text{sen},m,j}$ for  $j \in \mathcal{J}$. At the same time, the final row of the network output is directed into another FC layer, responsible for constructing a vector corresponding to the analog beamforming matrix $\textbf{F}_{m}$. Before obtaining the final outputs $\textbf{w}_{\text{com},m,i}$, $\textbf{w}_{\text{sen},m,j}$ and $\textbf{F}_{m}$, it is important to note that a normalization layer (NL) is employed is applied to satisfy the constant modulus constraint C2. 
\subsection{Train and Inference}
During the training phase, the GNN optimizes its weight and bias parameters using an unsupervised approach. The optimization objective is guided by a loss function, which is formulated as:
\begin{align}
\mathcal{L}= \frac{\sum_{t=1}^T f_\text{obj}^{(t)}}{T} ,  \nonumber
\end{align}
where $f_\text{obj}^{(t)} = \Big[ \alpha_\text{com}\sum_{i \in \mathcal{I}} \text{log}(1+ \gamma_i)+  \alpha_\text{sen}\sum_{j \in \mathcal{J}} \text{log}(1+ \eta\gamma_j) \Big]^{(t)}$ represents the WSCSC of the network for the $t$-th training sample, while $T$ denotes the total number of samples. During optimization, stochastic gradient descent (SGD) gradually minimizes the loss function, driving the objective value closer to its optimum. Once convergence is reached, the GNN successfully internalizes and models the complex dependencies between users and sensing targets. \par
Although training a GNN requires considerable computational effort, this process is carried out offline and therefore does not interfere with the system’s real-time operation. In contrast, the efficiency of the online inference stage is critical for assessing the feasibility of deploying the model in a MIMO cell-free ISAC network. To analyze this, we consider a single GNN, noting that in a multi-GNN architecture, each GNN executes inference independently. The first layer of the network is realized through an MLP, followed by two graph convolutional layers, each composed of two MLP blocks with nonlinear activation. The final outputs are generated by two FC layers. Consequently, the overall computational burden remains modest, with its exact level influenced by the quantization precision of the weight and bias parameters.
\section{EXTENSION}
This section extends the analysis to scenarios involving users and targets under imperfect CSI conditions. Specifically, the channel matrix $\tilde{\textbf{H}}_{\text{com},m,i}$ and $\tilde{\textbf{H}}_{\text{sen},m,j}$ can be expressed as 
\begin{align}
\tilde{\textbf{H}}_{\text{com},m,i} &= \bar{\textbf{H}}_{\text{com},m,i}+\hat{\textbf{H}} _{\text{com},m,i}, \nonumber \\
\tilde{\textbf{H}}_{\text{sen},m,j} &= \bar{\textbf{H}}_{\text{sen},m,j} + \hat{\textbf{H}}_{\text{sen},m,j}, \nonumber
\end{align}
where $\bar{\textbf{H}}_{\text{com},m,i}$ and  $\bar{\textbf{H}}_{\text{sen},m,j}$ denote the estimated CSI of the communication and cascaded sensing channels, respectively.  $\hat{\textbf{H}}_{\text{com},m,i}$ and $\hat{\textbf{H}}_{\text{sen},m,j}$ represent the channel errors of the communication and cascaded sensing channels, respectively. Following convention, we consider two channel error models: deterministic and stochastic. For the deterministic error model, the channel uncertainty is constrained within a bounded region, mathematically
expressed as
\begin{align}
| \hat{\textbf{H}}_{\text{com},m,i} (n_\text{u}, n_\text{t}) | \leq \epsilon_\text{com}, ~
| \hat{\textbf{H}}_{\text{sen},m,j} (n_\text{r}, n_\text{t}) | \leq \epsilon_\text{sen}, \nonumber
\end{align}
where thresholds $\epsilon_\text{com}$ and $\epsilon_\text{sen}$ denote the error bounds for communication and sensing, respectively. $n_\text{u}\in \mathcal{N_\text{u}}$, $n_\text{t}\in \mathcal{N_\text{t}}$ and $n_\text{r}\in \mathcal{N_\text{r}}$ with $\mathcal{N_\text{u}}$ and $\mathcal{N_\text{r}}$ being the antenna sets at each user and the radar receiver, respectively. The pairs $(n_\text{u}, n_\text{t})$ and $(n_\text{r}, n_\text{t})$ indicate the row and column indices of the elements in matrices $\hat{\textbf{H}}_{\text{com},m,i}(n_\text{u}, n_\text{t})$ and $\hat{\textbf{H}}_{\text{sen},m,j}(n_\text{r}, n_\text{t})$, respectively. For the stochastic error model, the channel error is assumed to follow a complex Gaussian distribution for simplicity. Mathematically, the channel error is given by
\begin{align}
\hat{\textbf{H}}_{\text{com},m,i} (n_\text{u}, n_\text{t}) & \sim \mathcal{CN} (0, \sigma_\text{com}^2), \nonumber \\
\hat{\textbf{H}}_{\text{sen},m,j}  (n_\text{r}, n_\text{t}) & \sim \mathcal{CN} (0,  \sigma_\text{sen}^2), \nonumber
\end{align}
where $\sigma_\text{com}^2$ and $\sigma_\text{sen}^2$ are the variances of the channel error for communication and sensing, respectively. \par
Based on this, the received signal at the $i$-th user and the radar-received signal from the $j$-th target are respectively given by \eqref{6} and \eqref{7}. Through the normalized receiving beamforming vector $\tilde{\boldsymbol{\textbf{u}}}_{j}$ at the radar receiver for the echo signal from the $j$-th target, the output is given by \eqref{8}. Then, the SINR of the $i$-th user and the radar receiver from the $j$-th target are respectively given by \eqref{9} and \eqref{10}.  
\begin{figure*}[b]
\hrulefill
\setcounter{equation}{5}
\begin{align}
\tilde{\textbf{y}}_{i}&=  \sum_{m \in \mathcal{M}} \bar{\textbf{H}}_{\text{com},m, i}\textbf{F}_m \textbf{w}_{\text{com},m,i} x_i +  \sum_{i^{\prime} \in \mathcal{I}, i^{\prime}  \neq i } \sum_{m \in \mathcal{M}} \bar{\textbf{H}}_{\text{com},m, i}\textbf{F}_m \textbf{w}_{\text{com},m,i'} x_{i'} +  \sum_{j \in \mathcal{J}} \sum_{m \in \mathcal{M}}  \bar{\textbf{H}}_{\text{com},m, i}\textbf{F}_m \textbf{w}_{\text{sen},m, j}\nonumber \\&~~ +\sum_{m \in \mathcal{M}} \hat{\textbf{H}}_{\text{com},m, i}\textbf{F}_m \textbf{w}_{\text{com},m,i} x_i +  \sum_{i^{\prime} \in \mathcal{I}, i^{\prime}  \neq i } \sum_{m \in \mathcal{M}} \hat{\textbf{H}}_{\text{com},m, i}\textbf{F}_m \textbf{w}_{\text{com},m,i'} x_{i'} +  \sum_{j \in \mathcal{J}} \sum_{m \in \mathcal{M}}  \hat{\textbf{H}}_{\text{com},m, i}\textbf{F}_m \textbf{w}_{\text{sen},m, j}  +\textbf{n}_{i},  \label{6} \\
\tilde{\textbf{y}}_{j}&= \sum_{m \in \mathcal{M}} \bar{\textbf{H}}_{\text{sen},m, j}\textbf{F}_m \textbf{w}_{\text{sen},m,j}+\sum_{j^{\prime} \in \mathcal{J}, j^{\prime}  \neq j } \sum_{m \in \mathcal{M}}\bar{\textbf{H}}_{\text{sen},m, j}\textbf{F}_m \textbf{w}_{\text{sen},m,j^{\prime}}+\sum_{i \in \mathcal{I} } \sum_{m \in \mathcal{M}}\bar{\textbf{H}}_{\text{sen},m, j}\textbf{F}_m \textbf{w}_{\text{com},m,i}x_i \nonumber \\ &~~ +\sum_{m \in \mathcal{M}} \hat{\textbf{H}}_{\text{sen},m, j}\textbf{F}_m \textbf{w}_{\text{sen},m,j}+\sum_{j^{\prime} \in \mathcal{J}, j^{\prime}  \neq j } \sum_{m \in \mathcal{M}}\hat{\textbf{H}}_{\text{sen},m, j}\textbf{F}_m \textbf{w}_{\text{sen},m,j^{\prime}}+\sum_{i \in \mathcal{I} } \sum_{m \in \mathcal{M}}\hat{\textbf{H}}_{\text{sen},m, j}\textbf{F}_m \textbf{w}_{\text{com},m,i}x_i +\textbf{n}_{j},\label{7} \\
\tilde{\bar{\textbf{y}}}_{j}&=  \sum_{m \in \mathcal{M}} \tilde{\boldsymbol{\textbf{u}}}_{j} \bar{\textbf{H}}_{\text{sen},m, j}\textbf{F}_m \textbf{w}_{\text{sen},m,j}+\sum_{j^{\prime} \in \mathcal{J}, j^{\prime}  \neq j } \sum_{m \in \mathcal{M}}\tilde{\boldsymbol{\textbf{u}}}_{j}\bar{\textbf{H}}_{\text{sen},m, j}\textbf{F}_m \textbf{w}_{\text{sen},m,j^{\prime}}+\sum_{i \in \mathcal{I} } \sum_{m \in \mathcal{M}}\tilde{\boldsymbol{\textbf{u}}}_{j}\bar{\textbf{H}}_{\text{sen},m, j}\textbf{F}_m \textbf{w}_{\text{com},m,i}x_i \nonumber \\ &~~ +\sum_{m \in \mathcal{M}} \tilde{\boldsymbol{\textbf{u}}}_{j}\hat{\textbf{H}}_{\text{sen},m, j}\textbf{F}_m \textbf{w}_{\text{sen},m,j}+\sum_{j^{\prime} \in \mathcal{J}, j^{\prime}  \neq j } \sum_{m \in \mathcal{M}}\tilde{\boldsymbol{\textbf{u}}}_{j}\hat{\textbf{H}}_{\text{sen},m, j}\textbf{F}_m \textbf{w}_{\text{sen},m,j^{\prime}}+\sum_{i \in \mathcal{I} } \sum_{m \in \mathcal{M}}\tilde{\boldsymbol{\textbf{u}}}_{j}\hat{\textbf{H}}_{\text{sen},m, j}\textbf{F}_m \textbf{w}_{\text{com},m,i}x_i +\textbf{n}_{j}, \label{8} \\
\tilde{\gamma}_{i} &= \left(\sum_{m \in \mathcal{M}}\bar{\textbf{H}}_{\text{com},m, i}\textbf{F}_m \textbf{w}_{\text{com},m,i}\right)^{H} \Bigg( \sigma_{i}^2 \textbf{I}_{i} +  \textbf{P}     \Bigg) ^{-1} \left(\sum_{m \in \mathcal{M}}\bar{\textbf{H}}_{\text{com},m, i}\textbf{F}_{m}\textbf{w}_{\text{com},m,i} \right), \label{9} \\
\textbf{P} &= ( \sum_{m \in \mathcal{M}}\hat{\textbf{H}}_{\text{com},m, i}\textbf{F}_m \textbf{w}_{\text{com},m,i})(\sum_{m \in \mathcal{M}}\hat{\textbf{H}}_{\text{com},m, i} \textbf{F}_{m}\textbf{w}_{\text{com},m,i})^H  \nonumber \\& ~~ +\sum_{i^{\prime} \in \mathcal{I}, i^{\prime}  \neq i }  ( \sum_{m \in \mathcal{M}}\bar{\textbf{H}}_{\text{com},m, i}\textbf{F}_m \textbf{w}_{\text{com},m,i'})(\sum_{m \in \mathcal{M}}\bar{\textbf{H}}_{\text{com},m, i} \textbf{F}_{m}\textbf{w}_{\text{com},m,i'})^H   \nonumber \\&~~  +\sum_{i^{\prime} \in \mathcal{I}, i^{\prime}  \neq i }  ( \sum_{m \in \mathcal{M}}\hat{\textbf{H}}_{\text{com},m, i}\textbf{F}_m \textbf{w}_{\text{com},m,i'})(\sum_{m \in \mathcal{M}}\hat{\textbf{H}}_{\text{com},m, i} \textbf{F}_{m}\textbf{w}_{\text{com},m,i'})^H   \nonumber \\
&~~ +\sum_{j \in \mathcal{J}}(\sum_{m \in \mathcal{M}} \bar{\textbf{H}}_{\text{com},m, i}\textbf{F}_{m} \textbf{w}_{\text{sen},m, j})( \sum_{m \in \mathcal{M}}\bar{\textbf{H}}_{\text{com},m, i}\textbf{F}_{m} \textbf{w}_{\text{sen},m, j})^H \nonumber \\ &~~  +\sum_{j \in \mathcal{J}}(\sum_{m \in \mathcal{M}} \hat{\textbf{H}}_{\text{com},m, i}\textbf{F}_{m} \textbf{w}_{\text{sen},m, j})( \sum_{m \in \mathcal{M}}\hat{\textbf{H}}_{\text{com},m, i}\textbf{F}_{m} \textbf{w}_{\text{sen},m, j})^H   \nonumber \\
\tilde{\gamma}_{j} &= \frac{|\sum_{m \in \mathcal{M}}\tilde{\boldsymbol{\textbf{u}}}_{j}\bar{\textbf{H}}_{\text{sen},m, j}\textbf{F}_{m}\textbf{w}_{\text{sen},m,j}|^2}{\sigma_{j}^2 |\boldsymbol{\textbf{u}}_{j}|^2 +  E},  \label{10} \\
E& = |\sum_{m \in \mathcal{M}}\tilde{\boldsymbol{\textbf{u}}}_{j}\hat{\textbf{H}}_{\text{sen},m, j}\textbf{F}_{m}\textbf{w}_{\text{sen},m,j}|^2+\sum_{j^{\prime} \in \mathcal{J}, j^{\prime}  \neq j } |\sum_{m \in \mathcal{M}}\tilde{\boldsymbol{\textbf{u}}}_{j}\bar{\textbf{H}}_{\text{sen},m,j}\textbf{F}_{m}\textbf{w}_{\text{sen},m,j^{\prime}}|^{2} +  \sum_{i \in \mathcal{I} } |\sum_{m \in \mathcal{M}}\tilde{\boldsymbol{\textbf{u}}}_{j}\bar{\textbf{H}}_{\text{sen},m, j}\textbf{F}_{m}\textbf{w}_{\text{com},m,i}|^{2}  \nonumber \\
& ~~ +\sum_{j^{\prime} \in \mathcal{J}, j^{\prime}  \neq j } |\sum_{m \in \mathcal{M}}\tilde{\boldsymbol{\textbf{u}}}_{j}\hat{\textbf{H}}_{\text{sen},m,j}\textbf{F}_{m}\textbf{w}_{\text{sen},m,j^{\prime}}|^{2} +  \sum_{i \in \mathcal{I} } |\sum_{m \in \mathcal{M}}\tilde{\boldsymbol{\textbf{u}}}_{j}\hat{\textbf{H}}_{\text{sen},m, j}\textbf{F}_{m}\textbf{w}_{\text{com},m,i}|^{2}.  \nonumber
\end{align}
\vspace*{0pt}
\end{figure*}
Therefore, the multi-objective optimization problem under imperfect CSI conditions is reformulated as\par
\begin{align*}
(\text{P}2)~~ \mathop{\text{max}}\limits_{\textbf{F}_m, \textbf{w}_{\text{com},m,i},\textbf{w}_{\text{sen},m,j}}      & \alpha_\text{com} \mathbb{E}_{\hat{\textbf{H}} _{\text{com},m,i}} \left[ \sum_{i \in \mathcal{I}} \text{log}(1+\tilde{\gamma}_i) \right] \\
& +  \alpha_\text{sen} \mathbb{E}_{\hat{\textbf{H}} _{\text{sen},m,j}} \left[ \sum_{j \in \mathcal{J}} \text{log}(1+\tilde{\gamma}_j)\right],\\
  \text{s.t.}  \quad  & \text{C1} ~ \text{and}~ \text{C2}. 
\end{align*}
It is evident that the proposed GNN-based optimization framework remains applicable with two main modifications: first, the GNN takes the estimated CSI instead of the perfect CSI as input; second, the loss function is reformulated.
%
%
\section{FPGA-based GNN Accelerator}
This section introduces the design of an FPGA-based accelerator intended to mitigate the latency associated with GNN inference. The discussion first details the microarchitecture of the proposed accelerator. Subsequently, the data transfer between on-chip and off-chip memory is described. Finally, the computation process is presented.
\subsection{Microarchitecture}
As depicted in Fig. 3, the proposed FPGA-based GNN accelerator is architecturally organized into three primary modules: the computation engine module, the memory module, and the control module. \par
\begin{figure}
\centering
\includegraphics[width= 3.0 in]{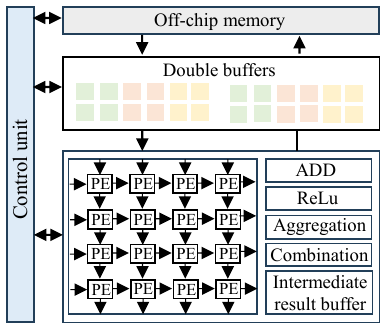}
\caption{Microarchitectural design of the FPGA-based GNN accelerator.}
\label{Fig3}
\end{figure}
\subsubsection{Computation Engine} All layers in the GNN model are implemented as FC layers, whose operations are dominated by large matrix multiplications. Accordingly, the computation engine is primarily built upon multiple systolic arrays (SAs) as its core processing units. The SA represents a dedicated hardware architecture characterized by a regular grid of processing elements (PEs), where each PE performs partial product computations, accumulates intermediate results, and transmits data to adjacent units in a synchronized, pipeline-like fashion. Such an architecture enables highly efficient parallel computation, rendering systolic arrays especially advantageous for workloads characterized by repetitive operations, including large-scale matrix computations. 
Besides the SAs, the computation engine also incorporates additional functional modules, such as the rectified linear unit (ReLU) and components responsible for add, aggregation and feature combination tasks.\par
\subsubsection{Memory} In terms of memory operations, a ping-pong style double buffering strategy is employed to allow data computation and data transfer to proceed in parallel. This approach enhances system throughput by reducing idle cycles. Moreover, the same buffering mechanism preserves the outputs of earlier layers so that they can be immediately reused as inputs to the following layers. Such reuse enables layer fusion, where multiple consecutive layers are executed as a single combined block. By substantially decreasing the number of intermediate data exchanges between off-chip and on-chip memory, both latency and the additional energy consumption resulting from frequent data transfers are reduced. \par
\subsubsection{Control Unit} The control unit is designed around a finite state machine (FSM), which governs the sequence of computational operations and the management of memory addresses. By employing an FSM, the controller can efficiently schedule computation tasks and data accesses, ensuring that operations follow the correct sequence and that the required data is available on demand. By adopting this structured approach, control logic design is streamlined, and the system’s efficiency and reliability are further improved.
\subsection{Data Transfer between On-Chip and Off-Chip Memory} 
Based on the roofline model \cite{Zhang2023GraphAGILE}, the inference latency of an FPGA is constrained not only by its computational resources but also by the data transfer between on-chip and off-chip memory. Since the target GNN contains a large number of parameters, including weights and biases, the primary factor affecting the accelerator’s inference speed is the I/O bandwidth needed for off-chip memory access, making memory operations the main performance bottleneck. To mitigate this limitation, four key optimization strategies are applied, including quantization technique, loop tiling technique, double buffering technique, as well as layer fusion technique. \par
Quantization technique refers to the process of reducing data bit-width, usually by mapping high-precision values (e.g., floating-point) to lower-precision formats such as fixed-point or integer. This method effectively lowers data transfer latency between on-chip and off-chip memory and improves computation efficiency on resource-constrained devices. However, it may also lead to some loss in model accuracy, requiring a careful balance between performance and precision. Loop tiling is a method for breaking down large loops into smaller segments, known as tiles. In an FPGA-based accelerator, each tile is processed individually, with tiles fed sequentially into the accelerator. The processing of each tile must be completed before the next tile begins, effectively splitting the loops into on-chip and off-chip components and improving on-chip data throughput. \par
\subsection{Computation Flow} The operation of the FPGA-based GNN accelerator is divided into three primary stages. Initially, the weight and bias parameters of the GNN are transferred from off-chip memory to on-chip memory using a double buffering mechanism. Next, the accelerator performs the necessary calculations to produce the analog precoding matrices and digital beamforming vectors for the ISAC network. In this stage, the dual-layer graph convolution operations of the GNN are reorganized to reduce the number of MLP layer executions, ensuring compatibility with the dimensions of the designed SA. Finally, the computed results are written back to off-chip memory. All computations in this stage are carried out entirely within on-chip resources, eliminating the need for external memory accesses during processing.
\section{Simulation and Experimental Results}
In this section, we first assess the communication and sensing performance of the proposed GNN-based method through numerical simulations, followed by presenting the experimental results on the computing performance of the FPGA-based accelerator.
\subsection{Communication Performance}
\begin{figure}
\centering
\includegraphics[width= 3.2 in]{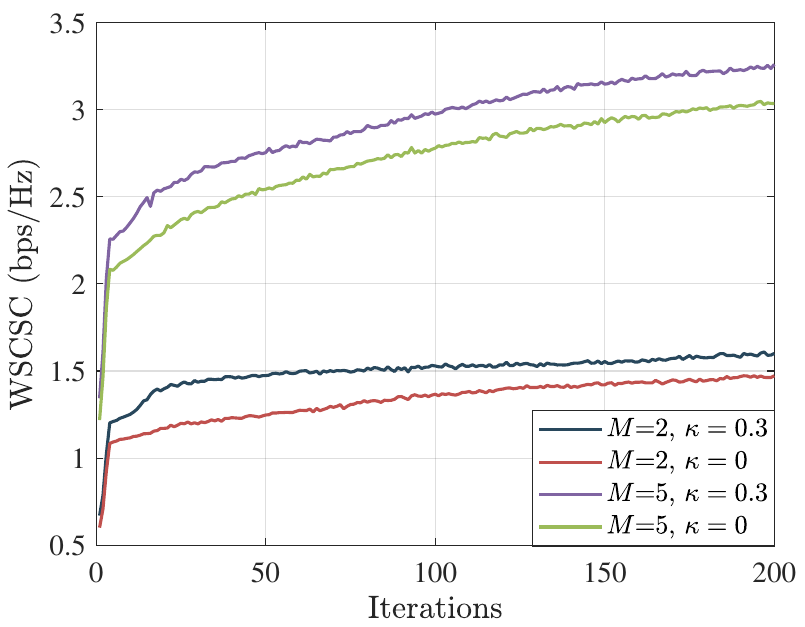}
\caption{WSCSC \emph{vs.} iteration number.}
\label{Fig4}
\end{figure}
\begin{figure}
\centering
\includegraphics[width= 3.2 in]{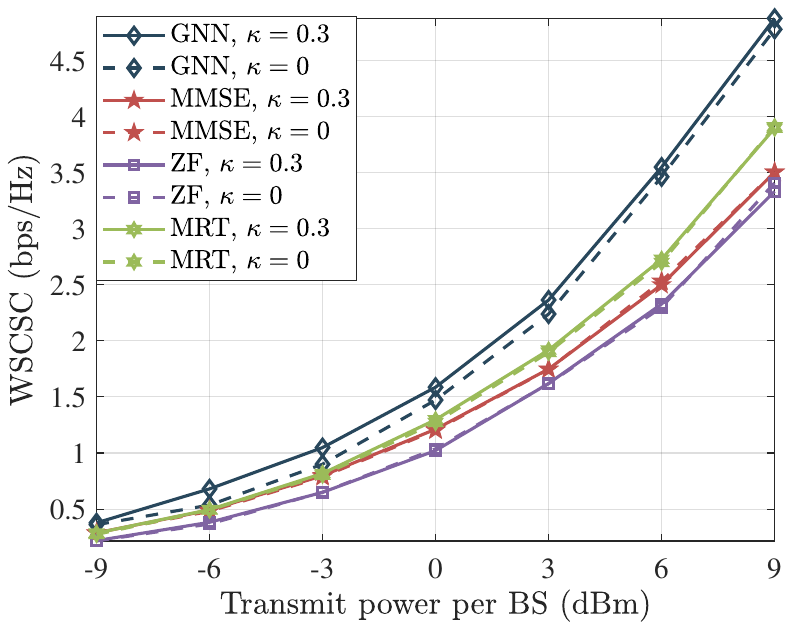}
\caption{WSCSC \emph{vs.} transmit power per BS.}
\label{Fig5}
\end{figure}
Simulations are carried out in this section to analyze the communication and sensing performance within the MIMO cell-free ISAC network. To evaluate the effectiveness of the proposed GNN-based method, the simulations consider multiple benchmark schemes, including minimum mean square error (MMSE), zero forcing (ZF), and maximum ratio transmission (MRT).  
\begin{table}[t]
  \centering
\scriptsize
  \caption{Parameter setting.}
  \begin{tabular}{|c|c|c|}
    \hline
    Notation & Description & Value \\
    \hline
    $P$ & Transmit power of each BS & 0 dBm \\
    \hline
    $\text{PL}_0$ & Path
loss & -30 dB \\
    \hline
    $\sigma_{i}^{2}$ & Noise variance of communication & $10^{-9}$ dBm \\
    \hline
    $\sigma_{r}^{2}$ & Noise variance of sensing & $10^{-9}$ dBm \\
    \hline
    $d_0$ & Reference distance & 1 m \\
    \hline
    $M$ & Number of BSs & 2 \\
    \hline
    $I$ & Number of users & 2 \\
    \hline
    $J$ & Number of targets & 2 \\
    \hline
    $N$ & Transmit RF chains of each BS & 6 \\
    \hline
    $N_\text{t}$ & Number of antennas at each BS & 16 \\
    \hline  
    $N_\text{u}$ & Number of antennas at each user & 2 \\
    \hline 
    $N_\text{r}$ & Number of antennas at radar receiver & 4 \\
    \hline
    $\alpha_\text{com}$ & Weighting factor of communication & 0.5 \\
    \hline
    $\alpha_\text{sen}$ & Weighting factor of sensing & 0.5 \\
    \hline
    $\eta$ & Amplification factor & $5 \times  10^{6}$ \\
    \hline
    $\kappa$ & Rician factor & 0.3 \\
    \hline
  \end{tabular}
\end{table}
\par Table I summarizes the parameters used in the simulations as follows: transmit power per BS is 0 dBm; $M=2$ BSs, $I=2$ users, and $J=2$ targets are considered. Each BS is equipped with $N_\text{t}=8$ antennas and $N=6$ RF chains, while each user has $N_\text{u}=2$ antennas and the radar receiver has $N_\text{r}=4$ antennas. The weighting factors are set to $\alpha_\text{com}=\alpha_\text{sen}=0.5$, and the amplification factor is $\eta=5 \times 10^{6}$. Some parameters vary depending on the simulation figures. In the simulations, channel realizations are generated randomly following either a Rician distribution with a Rician factor of $\kappa = 0.3$ or a Rayleigh distribution when $\kappa = 0$. The path loss is modeled as $\text{PL} = \text{PL}_0 - 25 \lg \left( {d}/{d_0} \right) $ dB, where $\text{PL}0 = -30$ dB represents the path loss at the reference distance $d_0 = 1$ m, and $d$ denotes the transmission distance. The distances between each BS and all users and targets are randomly sampled from the interval $[20\text{ m}, 30\text{ m}]$. All noise variances are set as $\sigma_{i}^{2} = \sigma_\text{r}^{2} = 10^{-9}$ dBm.\par 
\begin{figure}
\centering
\includegraphics[width= 3.2 in]{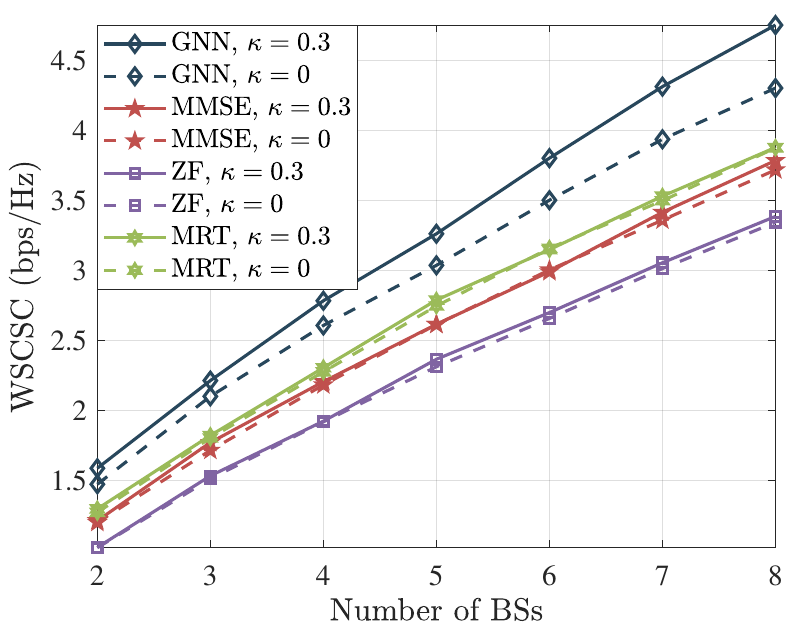}
\caption{WSCSC \emph{vs.} number of BSs.}
\label{Fig6}
\end{figure}
\begin{figure}
\centering
\includegraphics[width= 3.2 in]{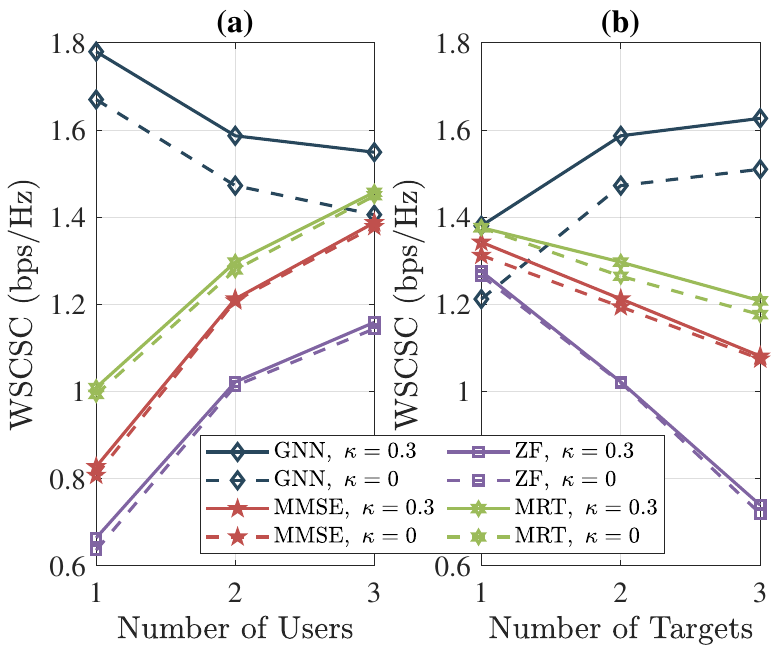}
\caption{WSCSC \emph{vs.} (a) the number of users and (b) the number of targets.}
\label{Fig7}
\end{figure}
\begin{figure}
\centering
\includegraphics[width= 3.2 in]{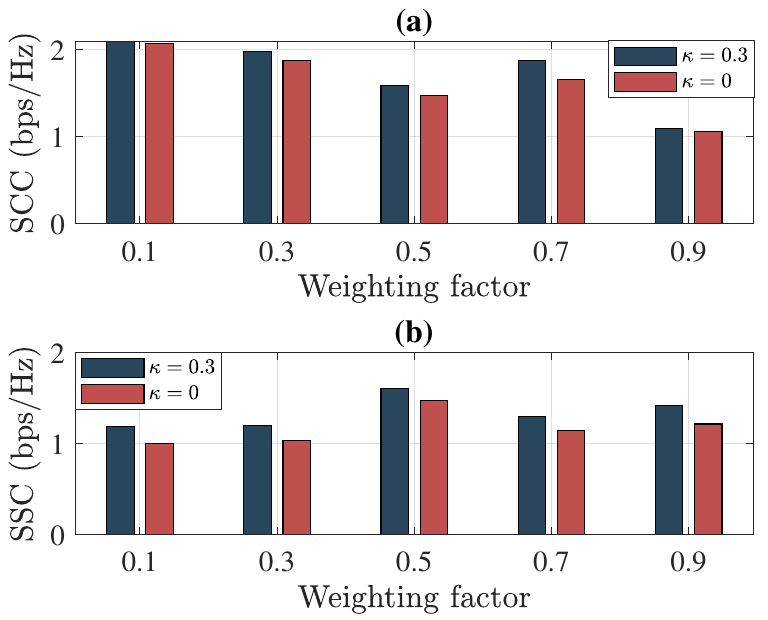}
\caption{(a) SCC \emph{vs.}  the weighting factor and (b) SSC \emph{vs.} the weighting factor.}
\label{Fig8}
\end{figure}
\begin{figure}
\centering
\includegraphics[width= 3.2 in]{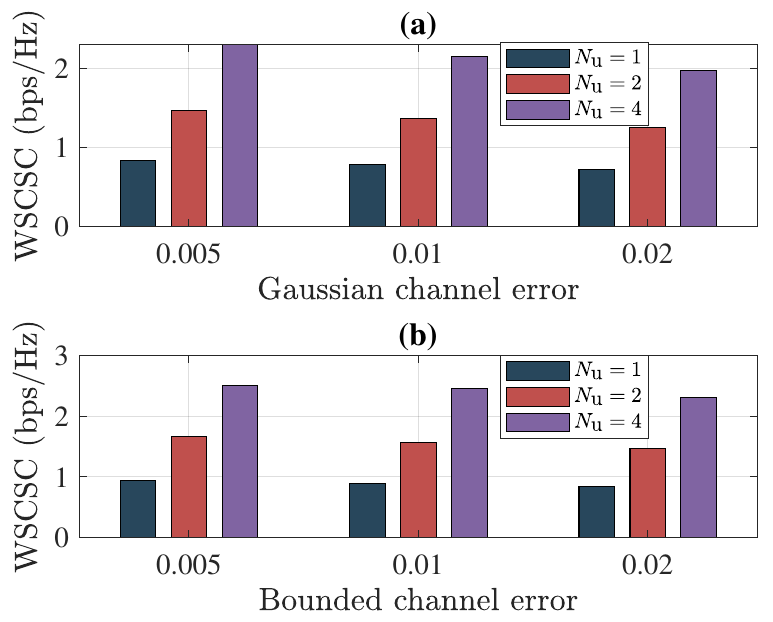}
\caption{WSCSC \emph{vs.} (a) the Gaussian channel error and (b) the bounded channel error.}
\label{Fig9}
\end{figure}
In the employed multi-GNN architecture, each GNN shares an identical structure. Specifically, a single GNN comprises two MLP layers, two graph convolution layers, and two FC layers. The detailed parameters for each layer are provided in Table II. Fig. \ref{Fig4} shows a typical example of the convergence characteristics of the proposed scheme. Across different values of $M$ and $\kappa$, the algorithm generally requires no more than 25 iterations to reach a satisfactory WSCSC.  \par
%
%
%
%
\begin{table*}[]
\tiny
\caption{GNN SETUP.}
\begin{tabular}{|cc|cccc|cccc|c|}
\hline
\multicolumn{2}{|c|}{MLP}               & \multicolumn{4}{c|}{GNN}             & \multicolumn{4}{c|}{GNN}               & \multirow{2}{*}{FC}  \\ \cline{1-10}
\multicolumn{1}{|c|}{FC}  & FC          & \multicolumn{2}{c|}{MLP}  & \multicolumn{2}{c|}{MLP}                 & \multicolumn{2}{c|}{MLP} & \multicolumn{2}{c|}{MLP} &              \\ \hline
\multicolumn{1}{|c|}{com $2N_\text{t} N_\text{u} \times 512$} & $512 \times 256$ & \multicolumn{1}{c|}{\multirow{2}{*}{$256 \times 256$}} & \multicolumn{1}{c|}{\multirow{2}{*}{$256 \times 256$}} & \multicolumn{1}{c|}{\multirow{2}{*}{$256 \times 256$}} & \multirow{2}{*}{$256 \times 256$} & \multicolumn{1}{c|}{\multirow{2}{*}{$256 \times 256$}} & \multicolumn{1}{c|}{\multirow{2}{*}{$256 \times 256$}} & \multicolumn{1}{c|}{\multirow{2}{*}{$256 \times 256$}} & \multirow{2}{*}{$256 \times 256$} & $256 \times 2N $ ~ $\textbf{w}_{m,i}$, $\textbf{w}_{m,j}$ \\ \cline{1-2} \cline{11-11} 
\multicolumn{1}{|c|}{sen $2N_\text{t} N_\text{r} \times 512$} & $512 \times 256$ & \multicolumn{1}{c|}{} & \multicolumn{1}{c|}{}   & \multicolumn{1}{c|}{}    &  & \multicolumn{1}{c|}{}  & \multicolumn{1}{c|}{}   & \multicolumn{1}{c|}{}   &      & $256 \times N_\text{t} N$  ~$\textbf{F}_{m}$  \\ \hline
\end{tabular}
\end{table*}
Fig. \ref{Fig5} shows the impact of the total transmit power $P$
on the WSCSC. Clearly, higher transmit power leads to an improvement in the WSCSC. Compared to MMSE, ZF and MRT, the proposed GNN-based optimization algorithm attains the highest WSCSC. In contrast, the ZF approach yields a relatively low WSCSC, while the MRT method, despite its simplicity and distributed nature, fails to achieve a satisfactory WSCSC. Fig. \ref{Fig6} depicts how the WSCSC varies with the number of BSs. The results indicate that WSCSC generally improves as the number of BSs increases, with the proposed GNN-based optimization algorithm consistently outperforming all benchmark methods. This improvement is attributed to the greater spatial degrees of freedom provided by additional BSs. However, the rate of improvement gradually diminishes as the number of BSs becomes larger. For $\kappa = 0$, the trend is similar to that for $\kappa = 0.3$, although the WSCSC values are slightly lower.\par
Fig. \ref{Fig7}(a) illustrates how the WSCSC varies with the number of users, considering the cases of $\kappa = 0.3$ and $\kappa = 0$. With more users are considered, the WSCSC achieved by the GNN-based optimization method decreases, whereas the performance of the benchmark schemes improves; moreover, as more BSs are considered, both the growth rate and the decay rate exhibit a gradual decline. Fig. \ref{Fig7}(b) shows how the WSCSC changes as the number of targets increases for the cases of $\kappa = 0.3$ and $\kappa = 0$. With an increasing number of targets, the WSCSC achieved by the GNN-based optimization method continues to rise, although the rate of growth gradually slows down, while the benchmark schemes exhibit a decline in performance at an almost constant rate.
\par
Fig. \ref{Fig8}(a) and Fig. \ref{Fig8}(b) depict the relationship between the sum communication capacity (SCC) and sum sensing capacity (SSC) with respect to $\alpha_\text{sen}$, for $\kappa = 0.3$ and $\kappa = 0$, respectively. For $\kappa = 0.3$, SCC decreases while SSC increases within the intervals $[0.1, 0.5]$ and $[0.7, 0.9]$. Conversely, when $\alpha_\text{sen}$ ranges from $0.5$ to $0.7$, SCC rises as SSC declines. A similar pattern is observed for $\kappa = 0$, though the corresponding values are slightly lower compared to the $\kappa = 0.3$ scenario. This indicates that the weighting factors have a strongly affect both communication and sensing performance.\par
Fig. \ref{Fig9}(a) and Fig. \ref{Fig9}(b) show how the number of user antennas and channel imperfections influence the WSCSC achieved by the GNN-based optimization method, considering Gaussian and bounded channel errors, respectively. The horizontal axis denotes the mean square error of channel estimation, with both error bounds, $\epsilon_\text{com}$ and $\epsilon_\text{sen}$, set at 0.05. It is evident that increasing the number of antennas significantly enhances the WSCSC, due to higher received power and greater spatial degrees of freedom. At the same time, the proposed GNN-based optimization approach exhibits notable robustness to channel inaccuracies, with only minor reductions in performance.
\subsection{Computing Performance}
This subsection evaluates the performance of the FPGA-based GNN accelerator. The implementation is carried out on a Xilinx Virtex-7 XC7V690T FFG1761-3 FPGA, with synthesis performed using Xilinx Vitis HLS 2022.2. To reduce resource utilization and power consumption, the accelerator adopts fixed-point arithmetic in place of floating-point operations. This approach also lowers the latency for transferring data between on-chip and off-chip memory. In the hardware design, 64 bits of the off-chip bandwidth are allocated to the accelerator, while the remaining bandwidth supports signal processing modules. Under a 10 ns clock period, the measured latency ranges from 432,638 to 658,873 cycles (equivalent to 4.326–6.588 ms). Furthermore, as the accelerator is primarily memory-bound, its latency can be reduced by increasing the data bit-width and lowering the quantization precision.
\section{Conclusions}
This paper presented a novel MIMO cell-free ISAC network architecture and proposed a GNN-based method for joint optimization of sensing and communication. Simulation results demonstrated that the algorithm converges efficiently and outperforms the MMSE, ZF, and MRT schemes in terms of overall communication and sensing performance. Moreover, experimental evaluation revealed that, with 8-bit fixed-point representation at a 10 ns clock period, the FPGA-based accelerator attains inference latency in the range of 3.863–5.883 ms.
%
%
%
%
%

%
%
%
%
%
\ifCLASSOPTIONcaptionsoff
  \newpage
\fi
\end{document}